# The role of grounded conducting pointy objects during Thunderstorm Ground Enhancements (TGEs)

Maribeth Stolzenburg and Thomas C. Marshall

Department of Physics and Astronomy, University of Mississippi, University, MS USA

Corresponding author: Maribeth Stolzenburg, mstolzen@phy.olemiss.edu

**Abstract**. A mechanism for increased fluxes of high energy particles detected in Thunderstorm Ground Enhancements (TGEs) is suggested. When the electric field $E_{ambient}$ at the ground beneath thunderstorms reaches modest magnitudes (6 -12 kV/m), corona discharge occurs from the tips of some grounded conducting pointy objects (GCPOs). Past theory and modeling of prolate semi-ellipsoidal GCPOs suggest the corona threshold value for free electron availability via seed electron detachment is $E = 67$ V/m/Pa at the tip. The field enhancement, $E_{tip}/E_{ambient}$, at the GCPO tip depends only on its length:tip radius of curvature ratio. Thus, $E_{tip}$ at all GCPOs is substantially enhanced by $E_{ambient}$, whether or not they are in corona. Cosmic ray electrons striking these pointy tips will gain MeV energies. These high energy electrons may be detected with instrumentation located nearby, or they may produce gamma rays via bremsstrahlung within the tips before being detected, thereby contributing to TGEs.

**1. Introduction**

   Thunderstorm Ground Enhancements (TGEs), as named by Chilingarian et al. (2010) and recently reviewed by Chilingarian et al. (2020), are increases in the flux of high energy particles (electrons, gamma rays, and neutrons) at the ground that last for several minutes to hours and occur during thunderstorms. The high-energy particles detected in short duration (several minutes long) TGEs have been attributed to runaway breakdown (Gurevich et al., 1992), also known as a relativistic runaway electron avalanche (RREA), in the very large electric field (E) suggested to exist between lower positive charge and mid-level negative charge regions (e.g., Chilingarian & Mkrytchyan, 2012) of a thundercloud. E measurements made at the ground during short duration TGEs show that they are often terminated by a nearby lightning flash (e.g., Khaerdinov et al., 2005; Tsuchiya et al., 2013; Chilingarian et al., 2017; Chum et al., 2020); airborne observations near and within thunderstorms have also found sharply increased X-ray intensities which cease at the time of a nearby lightning flash (e.g. McCarthy & Parks, 1985; Eack et al., 1996). Observed E values at the ground are usually (but not always) positive upward during the TGEs (e.g., Khaerdinov et al., 2005; Chum et al., 2020), thereby causing a downward force on electrons. Longer duration TGEs (1-2 hours) of increased high-energy particle fluxes have long been observed during thunderstorms (e.g. Shaw, 1967) and have been most clearly attributed to radon and daughter products (particularly $^{222}$Rn) by Bogomolov et al. (2018); this explanation was also suggested by Tsuchiya et al. (2009) and many other earlier researchers. The high-energy particle bursts of short duration TGEs usually occur during these longer duration TGE events.

   The particle detectors used by Chilingarian et al. (2010, 2020) to measure TGEs are located on Mt. Aragats at an altitude of 3.20 km msl. Bursts of strong TGEs have also been found using detectors on other mountain tops including in the Tien-Shan mountains at 3.30 km (Chubenko et al., 2000), on Mt. Fuji at 3.80 km (Torii et al., 2009), at 2.77 km on a mountain in Gifu prefecture (Tsuchiya et al., 2009), and at 2.60 km on Lomnický Štít (Chum et al., 2020). TGEs have also been observed at sea level during winter thunderstorms in Japan (e.g., Torii et al., 2004; Tsuchiya et al., 2013).

   The explanation given by Chilingarian and colleagues for short duration TGEs is dependent on the thunderstorm charge structure aloft. Results from many balloon-borne instruments through storms above Langmuir Laboratory (3.2 km msl) have found that both a lower positive charge and a main negative charge region are common in the thunderstorm convective region charge structure there (e.g. Stolzenburg et al., 1998). The predominant charge carriers of the lower positive charge region in convection are precipitation-sized particles (Bateman et al. 1999; Marshall & Winn 1982; Marshall & Marsh 1993; Marsh & Marshall 1993; Stolzenburg & Marshall 1998). Positive E values of 26-186 kV/m have been observed at altitudes of 4-6 km msl, between the lower positive and main negative charge regions (e.g., Stolzenburg et al., 1998; Marshall et al., 2005; Stolzenburg & Marshall, 2009). Although no in situ observations of the cloud charge structure or in-cloud E during TGEs have been made, there are reasons to expect that a substantial positive charge region and large positive E would exist in the thunderclouds above the mountain peaks where TGEs have been observed. At sea level along the coast of Japan mature winter thunderstorms also have a charge structure that includes a main negative charge and sometimes a lower positive charge region, but at lower altitudes, with maximum positive E inferred at about 2 km and cloud base at 0.2 -0.3 km (Kitagawa & Michimoto, 1994).

In this article, we present an additional mechanism that may contribute to the increased fluxes of high energy particles detected in TGEs. Our proposed mechanism relies on the basic fact that modest ambient E values of 5-12 kV/m occur at the surface beneath thunderstorm clouds. These modest ambient E values are known to result in point discharge, or corona, from grounded, conducting pointy objects at the surface. For example, at Lomnický Štít (High Tatras, 2.63 km, Slovak Republic) Chum et al. (2020) have reported that "The staff at the observatory located at the top of LS [Lomnický Štít] often observes St. Elmo's fire (corona discharges) during thunderstorms." Thus, at the tips of grounded, conducting pointy objects (GCPOs), the necessary very large E conditions (> 1 MV/m) exist and can produce electrons and photons with energies of 1-15 MeV; these products could be detected with instrumentation nearby. We suggest that this GCPO mechanism deserves further study by those with the suitable detailed observations in order to establish or eliminate its relevance to understanding TGE occurrence and variability.

**2. Point Discharge**

Herein we use the 'physics' definition of the vector E, with positive E oriented in the direction of the force it exerts on a positive charge. We only need to consider the vertical component of E, and we define positive E as upward. In addition, for the purposes herein we use the terms *point discharge* and *corona* (or corona discharge) interchangeably; there are technical differences between these in some literature, where the latter (also referred to as 'brush discharge') is more audible, more luminous, and occurs at greater field strengths.

Measuring E at the ground beneath a thunderstorm is a very useful way of studying storm electrification and evolution. However, surface E data from a single site do not provide unambiguous information about the location or magnitude of the charge(s) in the storm above, only the dominant charge polarity overhead at a given time. Furthermore, it has long been known that the interpretation of E measurements made at the ground is complicated by point discharge or corona (discharge) occurring from nearby structures and vegetation. In his studies of possible sources of electric charge in clouds, Wilson (1921) noted the importance of ions due to point discharges "from earth-connected conductors, such as the leaves of trees or even the tips of blades of grass, under the action of the intense electric field" beneath a thundercloud. According to the Glossary of Meteorology (AMS, 2021) "Point discharge is a major process of charge transfer between electrified clouds and the earth, and [it] is a leading item in the charge balance of the global electrical circuit."

Standler and Winn (1979) made extensive measurements of how point discharge both impacts and is affected by E at the ground. Standler and Winn [1979] stated that the "strength of the electric field at the extremities of sharp, grounded, elevated objects can be many hundred times that of the ambient field." Standler and Winn [1979] determined that the threshold ambient E needed to start point discharge was approximately 3 kV/m under Florida thunderstorms (near sea level, flat terrain) and was 5 kV/m at Langmuir Laboratory in New Mexico (3.2 km, on a mountain ridge); they suggested that the corona threshold in Florida was smaller because there are more corona points due to "dense bushes and tall, lush grass" versus the "relatively barren mountain ridge" in New Mexico.

The effects of point discharge are regularly observed near the surface below thunderclouds. By comparing E at the ground to E at a tethered balloon, Standler and Winn (1979) showed that corona ions emitted by point discharge were present up to at least 300 m above ground. Soula and Chauzy (1991) followed the evolution of the corona layer above the ground at sea level in Florida using five electric field meters at different altitudes on a tethered balloon. Soula and Chauzy (1991) found "that a large proportion of corona ions were carried up to several hundreds of meters above ground." Using E measurements with a rocket-borne sensor, Marshall et al. (1995) observed a shallow positive corona charge layer above ground at Langmuir Laboratory.

Just after launch, the lowest altitude of measured E was 10.3 kV/m, and it increased to 29.6 kV/m in the first 60 m above ground, indicating a corona layer charge density of 2.8 nC/m$^3$. Similar examples of distinct corona layers have been observed in many E soundings. Standler and Winn (1979) also observed that above the corona layer, E could be 2 – 6 times greater than at the ground. In Florida, Soula and Chauzy (1991) observed E values as large as 65 kV/m at 603 m, while the surface E "did not exceed 5 kV/m" during an active thunderstorm overhead.

Thus, once E beneath a cloud reaches the ambient threshold for point discharge, the charges produced by point discharge partially shield the surface E measurement from the dominant cloud charge above. In these conditions, the surface E measurement normally saturates at a magnitude close to the threshold for producing corona. For example, E magnitudes at the surface beneath thunderstorms near the Atlantic coast in central Florida rarely exceed 8 kV/m (e.g., Jacobson & Krider, 1976; Livingston & Krider, 1978), while at Langmuir Laboratory the E magnitude at the surface is typically less than 12 kV/m (e.g., Standler and Winn, 1979; Moore et al. 1986). Measured E magnitudes inside clouds can be larger, by a factor of 10 or more, than the E measured simultaneously at the surface (e.g., Marshall & Winn, 1982; Coleman et al., 2008; Marshall et al., 2009; Stolzenburg et al., 2015).

## 3. The Grounded Conducting Pointy Object (GCPO) mechanism

When corona is occurring at the ground below a thunderstorm, there is a large electric field at the tip ($E_{tip}$) of all GCPOs. This large E at *each* tip may give any cosmic ray secondary electrons that strike the tip energies up to ~15 MeV. Below we explain that the large $E_{tip}$ needed for corona depends on three parameters: the radius of curvature of the tip, the length of the GCPO, and the local atmospheric pressure. It is important to realize that a tip does not need to be in corona to have a large $E_{tip}$, rather the fact that *some* tips are in corona means that *all* tips of nearby GCPOs will have enhanced values of $E_{tip}$. Herein we focus on positive corona (positive charge on the tip in corona) rather than negative corona, since many of the minute-long TGE events occur during positive E and positive corona (e.g., Chum et al., 2020).

In studies of lightning rods, Moore et al. (2000, 2003) show how to calculate $E_{tip}$ of a GCPO. Note that corona is a pulsed discharge, as seen in Moore et al. (2003, their Figure 1). Each pulse in the corona discharge is an electron avalanche caused by $E_{tip}$. The electron avalanche needs one free electron to start, but free electrons are quite rare in the atmosphere because they quickly attach to $O_2$ or $H_2O$ molecules to make negative ions. However, for a sufficiently large $E_{tip}$, a negative $O_2$ or $H_2O$ ion will gain enough energy in one mean free path length to detach its extra electron upon collision with another air molecule, and then the electron avalanche can begin (Kip, 1938). Loeb (1935) found that the critical E needed to detach an electron from a negative $O_2$ ion was about 90 V/cm/torr or 67 V/m/Pa in SI units. Thus, at sea level the critical $E_{tip}$ to start corona is 6.79 MV/m. At 3.20 km altitude (69.2 kPa), the critical $E_{tip}$ is 4.63 MV/m.

Once the critical $E_{tip}$ is reached, free electrons will be liberated from many nearby negative molecular ions. These free electrons are then accelerated by $E_{tip}$ and acquire enough energy to liberate additional electrons from air molecules, thereby making an electron avalanche. The avalanche process stops when enough positive molecular ions surround the tip to reduce E below the critical value. The positive molecular ions will be driven away from the tip by its positive E; when they are far enough away, $E_{tip}$ will again have the critical magnitude to repeat the above process. (If there is any wind, the positive molecular ions will be swept away more quickly.) These repeating avalanches give the corona discharge its pulsing nature. In Figure 1b of Moore et al. (2003), the corona discharge rate was about 50 kHz, or 20 µs between pulses, and the pulse duration (duration of each electron avalanche) was approximately 2 µs.

As discussed above, during a thunderstorm with a positive ambient E at the surface, positive corona ions from the ground drift upward, thereby creating a layer of corona ions that reduce E at the surface to an ambient value ($E_{ambient}$) < 12 kV/m. Moore et al. (2000, 2003) imagined a corona point as a vertical, cylindrical, conducting rod with length *c* and upper radius of curvature *a*, sitting on the conducting Earth, and they modeled this rod as a prolate semiellipsoid. Moore et al. (2000) determined the critical field enhancement factor, $k_e$, caused by the rod is given by (from Moore et al., 2003):

$$k_e \equiv \frac{E_{tip}}{E_{ambient}} = \left[\eta_0 \left(\eta_0^2 - 1\right) \left(ArcCoth\, \eta_0 - \frac{1}{\eta_0}\right)\right]^{-1} \quad (1)$$

where

$$\eta_0 = \left(1 - \frac{a}{c}\right)^{-1/2} \quad (2)$$

Figure 1 shows $E_{tip}$ as a function of *c/a* for four example values of $E_{ambient}$ (6, 8, 10, and 12 kV/m) observed under thunderstorms at Langmuir Lab. The critical field values needed to start corona at 3.20 km (4.63 MV/m) and at sea level (6.79 MV/m) are indicated by horizontal guidelines in Figure 1. Twenty values of *c/a* between 250 and 5000 were used to represent the quasilinear form of $E_{tip}$ with *c/a* expressed in Eqs. 1 and 2.

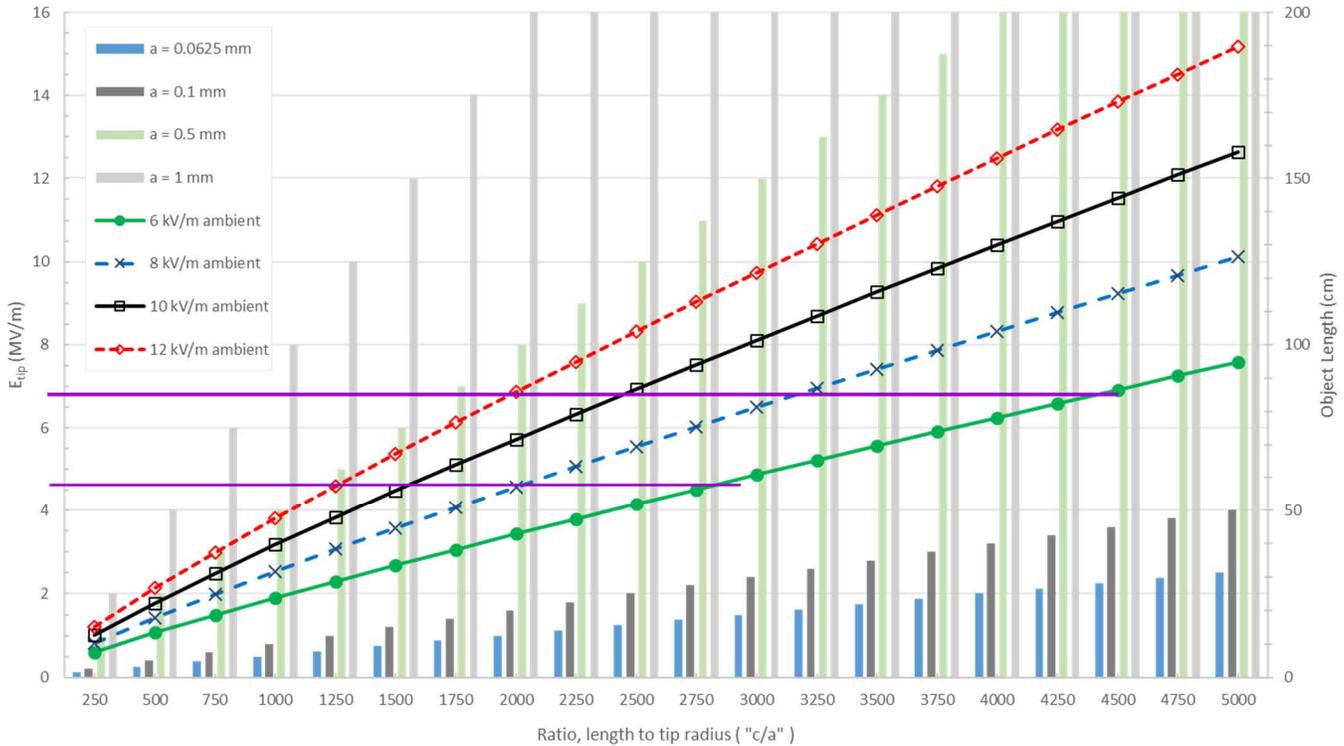

**Figure 1.** Enhanced field $E_{tip}$ at the tip of a Grounded Conducting Pointy Object as a function of the ratio of object length to radius of curvature ($c/a$), for four ambient E values. Horizontal lines at 4.63 MV/m and 6.79 MV/m indicate critical field needed for corona at ground levels of 3.20 km and 0 km altitude. Bars show object length $c$, in cm (right axis), for four values of $a$ at the different ratios. Lengths for larger ratios of the larger two radii go off-scale but are readily calculable.

Figure 1 also includes, also as a function of the ratio $c/a$, the object length $c$ for four example values of $a$: 0.0625, 0.1, 0.5, 1.0 mm. (We note here for reference that common, straight, 'Dressmaker' pins used in sewing have shafts of radius 0.25 mm or 0.35 mm and tip radii of curvature 0.0625-0.0635 mm, and a common pin head has approximately 1.27 mm radius.) Obviously, for the same $c/a$, sharper tips have much shorter lengths than blunt tips; sharp-tipped GCPOs are probably representative of grass, wildflowers, and various weeds (all of which are conductors, especially when wet from rain). Blunt tips generally require meter lengths for appreciable $E_{tip}$ values, and these GCPOs are probably representative of trees, metal posts, and metal frameworks (such as those used to support sensors). At most sites there are likely many other grounded, conducting objects that do not resemble vertical pointed rods but nonetheless should have substantial values of $E_{tip}$ at edges or points: e.g., metal roofing (and screw heads), metal siding on buildings (and sharp edges), and barbed-wire or other metal fencing.

Moore et al. (2000) also show that the enhanced field at the tip falls off rapidly above the tip, decreasing to 1% of $E_{tip}$ in a distance of $50a$ or about 3 – 50 mm for the values of $a$ used in Figure 1.

With the above background, we can now describe how GCPOs can give a cosmic ray secondary electron a large energy. Suppose such an electron moves downward directly toward the tip of a GCPO located at 3.20 km altitude when the surface electric field $E_{ambient}$ = +6 kV/m. If the $c/a$ of the object is 1250, then $E_{tip}$ = 2.30 MV/m (see Figure 1). This is less than the critical E of 4.62 MV/m needed for corona (at 3.20 km), so the tip is not emitting corona ions. However, as the electron moves downward to the GCPO tip, it will go from being in an E of 6 kV/m, to an E of 1% of $E_{tip}$ at $50a$ (3 – 50 mm) above the tip, and then to $E_{tip}$ = 2.3 MV/m just as it hits the point; hence this electron should gain 0.99 • 2.30 MeV or 2.28 MeV from the electric field in the last 3 – 50 mm before hitting the point. Because 3 – 50 mm is so short, the electron will lose an insignificant amount of energy due ionization of air molecules. With $c/a$ = 1250, for example (Figure 1): with $a$ = 0.0625 mm the length is 7.8 cm, while for $a$ = 0.5 mm the length is 62.5 cm. Furthermore, from Figure 1 we see that if $E_{amb}$ = 6 kV/m (a common and long-lived occurrence under thunderstorms at Langmuir Lab), then electrons hitting the tips of GCPOs gain energies in the range 0.6 MeV – 7.5 MeV; the larger of these values are large enough for the tips to be in corona (i.e., $E_{tip}$ > 4.62 MV/m). If $E_{amb}$ = 12 kV/m, then electrons hitting the tips of GCPOs gain energies in the range 1.2 MeV – 15.0 MeV.

If there are high energy particle detectors located nearby the GCPOs, it seems plausible that many of the high energy electrons (discussed in the previous paragraph) will not reach the detectors. However, there are at least two ways in which a significant number of these high energy electrons might be detected. First, as a high energy electron enters the GCPO that enhanced its energy, it is possible for it to interact with atoms of the object and produce bremsstrahlung gamma rays that could more easily reach a detector. Second, the GCPO might be the edge of a detector or a detector support, so that the electrons would be more likely to enter the detector.

For a full calculation of this process it will be necessary to know the number of GCPOs contributing to electron energy enhancement. We can make one estimate of the number of objects actively producing corona based on data in Standler and Winn (1979). They found one corona layer at Langmuir Lab developed quickly (~3 minutes, based on their Figure 4) when the corona current from short trees (1.4 m tall) was ~30 nA per tree. The surface E was 6.4 kV/m, the measured the corona layer thickness was 150 m, and the calculated charge density was 0.2 nC/m$^3$. We can reasonably take the horizontal area of the corona layer as 1 km$^2$, making a total charge in the layer of 0.03 C. It would take ~6000 trees emitting 30 nA each to produce a corona layer of 0.03 C in 3 minutes. If there are 6000 points in corona in the 1 km$^2$, then based on Figure 1, these 6000 points have $c/a > 2700$. One could reasonably expect that there would be many more points with $c/a < 2700$ (i.e., with shorter lengths and/or larger tip radii). Thus one might estimate of order $10^4$ - $10^5$ GCPOs in 1 km$^2$, all of which would be available to produce high energy electrons when the ambient surface E is greater than 5 kV/m. A rough estimate of how many cosmic ray electrons might be expected to hit the tip of a GCPO is as follows. To estimate the background cosmic ray flux at Mt Aragats, we used EXPACS (EXcel-based Program for calculating Atmospheric Cosmic-ray Spectrum; Sato, 2015; 2016) for a July day in 2017, and found 0.224 electrons/cm$^2$/s in the energy range of 0.01 – 9 MeV. This yields about 0.42 electrons per minute per tip within the cross-sectional area of one 'blunt' GCPO tip, with $a = 1$ mm. For a 'sharp' GCPO, with $a = 0.0625$ mm, the expected yield is 0.00164934 electrons per minute per tip. Thus, with perhaps $10^4$ - $10^5$ GCPOs or more per km$^2$, the proposed mechanism might account for 70-700 s$^{-1}$ high-energy radiation and particles across all energies available for detection at Mt. Aragats.

At the Japan coastal site (0 km msl) of Tsuchiya et al. (2013), the background cosmic ray flux on a January day in 2017 determined from EXPACS is about a factor of ten lower, about 3.05 x 10$^{-2}$ cm$^{-2}$ s$^{-1}$ and only the lower energy range of 0.01 eV to 10 keV is present. This flux would give 0.05746 electrons per minute arriving within the cross-sectional area of our blunt GCPO tip (with $a = 1$ mm). Thus, an electron might be expected to arrive every 17.4 minutes, at every similar point. In the ambient E of 6 kV/m beneath a thunderstorm, these arriving electrons may interact with the E at each tip to possibly cause bremsstrahlung and detectable particles with energies of 1 MeV, or greater (3-7 MeV) if the ambient E is 12 kV/m (Figure 1). If there were $10^4$ - $10^5$ similar and smaller GCPOs per km$^2$, the proposed mechanism might account for ~ 10 – 100 s$^{-1}$ high-energy radiation and particle flux across all energies at the Japan coastal site.

The fluxes estimated above could readily represent part of the enhancements detected as TGEs with long durations (30 minutes or more), as well as the shorter (1-10 minutes), more intense TGEs. A greater enhancement, and especially increases of higher energy detections (i.e, the so-called "hardening of the spectrum"), could be expected at any time when the ambient surface E was larger (10-12 kV/m), since all the $E_{tip}$ values and consequent electron energy gains would be larger in those conditions. Rapid, short-term increases of the ambient surface E are frequently observed in the few minutes leading up to a nearby lightning flash. The differences among the curves in Figure 1 for the different $E_{ambient}$ clearly show the effect larger values have on the $E_{tip}$, and $E_{tip}$ relates directly to the possible energy gain. In addition, a sudden cessation of this flux enhancement would occur whenever the ambient surface E was reduced suddenly (e.g., after a lightning flash) below a value at which none of the GCPOs would have significant $E_{tip}$ value. This direct field dependence of the GCPO mechanism can thus help explain the sudden 'termination' of TGEs when lightning occurs in the area which reduces the surface E.

Thus, we propose that the many GCPOs over a wide area beneath an active thunderstorm could account for some of the energetic particles detected as a TGE when the electric field at the ground is positive. In a modest positive E (6-12 kV/m), the many GCPOs can boost the energy spectrum of incoming cosmic ray electrons and can provide detectable gamma ray photons. As long as the surface E beneath a thunderstorm is positive and modestly large, there should be increased fluxes of high energy photons and electrons observed at nearby detectors. At Langmuir Laboratory, the surface E is at or above the 5 kV/m for most of a thunderstorm duration, typically 30 minutes to 2 hours. Thus, it seems plausible that the GCPO mechanism can contribute to the fluxes of high-energy particles detected in TGEs.

## 4. Additional support for the GCPO hypothesis

Chilingarian and Mkrtchyan (2012) and Chilingarian et al. (2019, 2020, and earlier works referenced therein) associate TGEs with very large magnitudes of E (or its negative, the potential gradient) measured near the surface. According to Chilingarian et al. (2019) "All TGEs are related to the large disturbances reaching -20kV/m." In each TGE case shown in the above articles, the measured peak E magnitudes range from approximately 18 kV/m to 36 kV/m. (Tables 1 and 2 in Chilingarian et al. (2019) list values ranging from -0.5 to -42 kV/m during 105 TGEs, with 64 cases >17 kV/m.) These E values at Mt. Aragats are very large, and they indicate to us that corona is likely occurring from many GCPOs in the surrounding area. Based on Eqs. 1 and 2, $E_{tip}$ values would be extremely large throughout the surrounding area: for example, using $c/a = 2000$ gives $E_{tip} = 11.44$ MV/m for the $E_{ambient}$ of 20 kV/m (i.e., twice that shown for $E_{ambient}$ of 10 kV/m in Figure 1). In fact, the E and count rate records in the three above-mentioned articles indicate that flux rates nearly always increase when the measured E exceeds ~5 to 8 kV/m in magnitude and nearly always subside when E magnitudes drop below 5 kV/m. This relationship provides nominal support for the GCPO hypothesis we present herein.

Since the E magnitudes given by Chilingarian and colleagues are substantially larger than those typically measured beneath thunderstorms, it should be ascertained if those measurements suffer from site error effects (such as rainfall, splashing, corona, or wind in nearby vegetation). Another possible problem with measuring surface E is that charge can be induced on the instrument itself by the thunderstorm field; unless it is accounted for by determining the 'form factor' (via site calibration) of the mill installation, this induced charge would be included in the measured surface E. Calibration with a known reference sensor that is flush-mounted in flat ground and away from nearby sources of space charge emission (pointy objects), as described in Rust and MacGorman (1988) and Antunes de Sá et al. (2020), would be highly worthwhile during thunderstorm conditions to ensure the accuracy of the E measurements during TGEs. However, even without a form factor correction, the E mill data from Mt. Aragats nicely show the thunderstorm evolution and the rapid growth of E during short, intense TGEs.

## 5. Summary

Thunderstorm Ground Enhancements (TGEs) of high-energy particles have been detected near sea level and on several mountaintops around the globe, as described above. Some of these studies have concluded that intense TGE enhancements (typically lasting only a few minutes) are caused by relativistic runaway electron avalanches (RREAs) associated with the large electric fields suggested to exist between the lower positive charge and main negative charge region in the thunderclouds (e.g., Chilingarian et al. 2020, 2021). As argued by Chilingarian and colleagues, Torii et al. (2011) and others, their conclusions require that the large E values exist within about 300 m above ground for detection of the TGEs at ground level. However, no measurements of intense TGEs include E data from within the parent thundercloud.

In situ measurements of electric field through many active thunderstorms are well described in the published literature. Some of these measurements have revealed E values sufficient for RREA (e.g., Marshall et al., 2005; Stolzenburg et al., 2007; Stolzenburg & Marshall, 2009). However, such large E magnitudes have not been detected at low altitude, below about 5 km, and not less than 2 km above ground. Thus, additional possible explanations for TGEs seem warranted. The possible explanation offered in this article we term the Grounded Conducting

Pointy Object (GCPO) mechanism.

The GCPO mechanism requires a large number of sufficiently pointy GCPOs, and we estimated $10^4 – 10^5$ per km$^2$ based on thunderstorm corona data of Standler and Winn (1979). The GCPO mechanism requires a large $E_{tip}$ for each sufficiently pointy GCPO, and Equation 1 gives values of 0.5 – 15 MV/m (shown in Figure 1) for modest ambient E values. Equation 1 also implies only modest $E_{ambient}$ values (5 – 12 kV/m) at the ground below a thunderstorm are required, and these values have routinely been measured under all thunderstorms. Thus, we assert that secondary cosmic ray electrons hitting the tips of sufficiently pointy GCPOs located under active thunderstorms should acquire energies of 0.5 – 15 MeV, and we further suggest that some of these high energy electrons might produce gamma rays via bremsstrahlung within the GCPO.

As described above, $E_{ambient}$ values of 5 – 12 kV/m can have durations of most of the mature stage of a thunderstorm, so the GCPO mechanism can easily contribute throughout long duration TGEs (10 – 80 minutes). In addition, $E_{ambient}$ often goes through the following sequence during a thunderstorm: decreases to a small value (1 – 2 kV/m) at the time of a nearby lightning flash, increases through ~ 2 – 4 minutes to a "saturated" (corona threshold) value of 5 – 8 kV/m, briefly increases further to ~ 10 – 12 kV/m through ~ 1 minute prior to another flash, then decreases again at the time of the next nearby lightning flash. Based on the GCPO mechanism, at the beginning of this sequence no enhancement of high energy particles would be detected, then during the increase in $E_{ambient}$, the number of high energy particles and their range of energies would increase as more and more GCPOs had $E_{tip}$ values > 1 MV/m, and then with the next lightning flash, the enhanced flux of high energy particles would cease. This sequence fits well with the description of "short duration TGEs" reviewed in the Introduction.

We note that the high energy particle detectors that record TGEs are also located on or near the ground beneath the thunderstorms, likely surrounded by GCPOs. The detectors may also experience $E_{ambient}$, so if the detectors are housed in metal boxes or held in a metal framework, then E at any sharp edges of boxes or metal framing might reach MV/m values, and cosmic ray secondary electrons might gain MeV energies at these edges. Thus, we propose that MeV electrons and gamma rays from the GCPO mechanism should be contributing to TGEs. We hope that this hypothesis will be examined via modeling and further observations.

**Acknowledgements.** No new data are used in this article. We appreciate the reviewers of a prior version of this manuscript for their comments which helped us improve it. We are grateful to Earle Williams for returning our attention to the matter of TGEs and for helpful discussions. We also thank Alexander Kostinskiy for earlier discussions alerting us to EXPACS capabilities. The EXPACS program for calculating the cosmic ray spectrum is available at http://phits.jaea.go.jp/expacs/ (8 Aug 2021). Work on this article was supported by U.S. National Science Foundation grant AGS-1745931.

**References**.

Wilson, C. T. R. (1921). Investigations on lightning discharges and on the electric field of thunderstorms. *Philosophical Transactions of the Royal Society of London. Series A, Containing Papers of a Mathematical Or Physical Character, 221*(582-593), 73-115. https://doi.org/10.1098/rsta.1921.0003